\pdfoutput=1
\documentclass[twocolumn,english,aps,superscriptaddress,prb,showpacs]{revtex4}

\usepackage{babel}
\usepackage{amsmath}
\usepackage{amssymb}
\usepackage{graphicx}

\begin{document}

\title{Berry phase investigation of spin-S ladders}

\author{Natalia Chepiga}
\affiliation{Institute of Theoretical Physics, Ecole Polytechnique F\'ed\'erale de Lausanne (EPFL), CH-1015 Lausanne, Switzerland}\author{Fr\'ed\'eric Michaud}
\affiliation{Institute of Theoretical Physics, Ecole Polytechnique F\'ed\'erale de Lausanne (EPFL), CH-1015 Lausanne, Switzerland}
\author{Fr\'ed\'eric Mila}
\affiliation{Institute of Theoretical Physics, Ecole Polytechnique F\'ed\'erale de Lausanne (EPFL), CH-1015 Lausanne, Switzerland}

\date{\today}
\begin{abstract} 
We investigate the properties of antiferromagnetic spin-S ladders with the help of local Berry phases defined 
by imposing a twist on one or a few local bonds. In gapped systems with time reversal symmetry, these Berry phases are 
quantized, hence able in principle to characterize different phases. In the case of a fully frustrated ladder where the total spin on 
a rung is a conserved quantity that changes abruptly upon increasing the rung coupling, we show that two Berry phases are 
relevant to detect such phase transitions: the rung Berry phase defined by imposing a twist on one rung coupling, and 
the twist Berry phase defined by twisting the boundary 
conditions along the legs. In the case of non-frustrated ladders, we have followed the fate of both Berry phases
when interpolating between standard ladders and dimerized spin chains, with the surprising conclusion that, at least
far enough from dimerized chains, they define different domains in parameter space. A careful investigation of
the spin gap and of edge states shows that a change of twist Berry phase is associated to a quantum phase transition at
which the bulk gap closes, and at which, with appropriate boundary conditions, edge states appear or disappear, while a
change of rung Berry phase is not necessarily associated to a quantum phase transition. The difference is particularly
acute for regular ladders, in which the twist Berry phase does not change at all upon increasing the rung coupling from 
zero to infinity while the rung Berry phase changes 2S times. By analogy with the fully frustrated ladder, these changes 
are interpreted as cross-overs between domains in which the rungs are in different states of total spin from 0 in the strong 
rung limit to 2S in the weak rung limit. This interpretation is further supported by the observation that these cross-overs turn into 
real phase transitions as a function of rung coupling if one rung is strongly ferromagnetic, or equivalently 
if one rung is replaced by a spin 2S impurity. 
\end{abstract}

\maketitle


\section{Introduction}
Topological aspects of matter have become one of the dominant themes in solid state physics. The characterization of quantum phases by topological invariants has been an extremely fruitful concept in the Quantum Hall Effect\cite{thouless,avron}, and it lies at the root of more recent developments such as topological insulators\cite{hasan_kane,qi_zhang}. A number of the central concepts have already been
discovered quite some time ago in the context of quantum magnetism. Indeed, in the modern language of topological matter, the
ground state of the spin-1 chain is in a topologically non-trivial phase characterized by a string order parameter\cite{PhysRevB.40.4709} 
and by spin-1/2 edge states\cite{kennedy}. In recent years, the investigation of the topological properties of other models of quantum 
magnetism, in particular spin-1/2 ladders, has been a very active field of research\cite{PhysRevLett.77.3443,sierra1,todo,PhysRevB.76.184428,PhysRevB.77.094415,1751-8121-41-48-485301,kim,poilblanc,chitov,Chen,pollmann}. 

A few years ago, Hatsugai proposed an alternative characterization of quantum magnets in terms of a Berry phase defined by 
twisting the XY components of the spin-spin interaction of one or several local bonds\cite{hatsugai,hatsugai2}. He showed that, if the
system has time-reversal symmetry, this Berry phase is quantized and can only take the values 0 or $\pi$ (mod. $2\pi$), 
and that some phase transitions can be characterized by a change of Berry phase. This is for instance the case of the dimerized
spin-S chains with $S>1/2$. As predicted a long time ago by Affleck and Haldane\cite{affleck_haldane}, they undergo upon increasing the dimerization
a series of 2S phase transitions at which, as shown by Hatsugai and collaborators for S=1 and 2,\cite{hirano_katsura_hatsugai} 
the Berry phase of a bond 
changes between $0$ and $\pi$. A simple explanation of these transitions is provided by the valence-bond singlet picture,
according to which they correspond to increases by 1 of the number of valence-bond singlets on the strong bonds up to 2S in the limit
of the fully dimerized chain.

One could naively expect that similar transitions occur in spin ladders upon increasing the rung coupling since the limit
of very strong rungs is similar. There is ample evidence however, at least for S=1/2 and S=1 ladders, that this is not the case. 
Spin ladders are defined by the Hamiltonian
 \begin{equation}
H^{\rm Ladder}=J_\parallel\sum_i\sum_{\alpha=1,2}{\bf S}_{i,\alpha}\cdot{\bf S}_{i+1,\alpha}+J_{\perp}\sum_i{\bf S}_{i,1}\cdot{\bf S}_{i,2}
\end{equation}
where $i$ is the rung index, $\alpha$ the leg one, and ${\bf S}_{i,\alpha}$ are spin S operators. To describe the whole 
parameter range, it is usual to introduce the parametrization $J_\parallel=J \cos \theta$ and $J_\perp=J\sin \theta$, 
and this convention will be used throughout.  Spin-1/2 ladders have 
been investigated in great detail over the last two decades\cite{dagotto_rice}. 
It has been shown early on that, for antiferromagnetic rung coupling, they are gapped if the number of legs is even
and gapless if it is odd,  and for the two-leg ladder, a consensus has emerged, based on numerical investigations and field theory arguments, 
that there is no quantum phase transition between the weak and strong rung coupling regimes.
The only phase transition that has been detected in spin-1/2 ladders with antiferromagnetic leg coupling takes place when the sign of the rung coupling changes from antiferromagnetic to ferromagnetic. At this transition, the gap closes, and the two topologically distinct singlet phases can be distinguished by the type of string-order parameter (even or odd) that exhibits long-range order\cite{nishiyama,kim_fath_solyom_scalapino,fath_legeza_solyom}. 
And for the spin-1 two-leg ladder \cite{todo_munehisa}, the evidence has imposed itself that the spin gap remains open all the way
from weak to strong rung coupling, definitely excluding the presence of a quantum phase transition. 

In this paper, one of our goals is to clarify the difference between spin ladders and dimerized spin chains 
 by a systematic investigation of different Berry phases in a model
that interpolates between them. As we shall see, the phase transitions of
the dimerized spin chains disappear before reaching the spin ladder geometry, in agreement with previous
results \cite{sierra1,PhysRevLett.77.3443,PhysRevB.76.184428,PhysRevB.77.094415,1751-8121-41-48-485301,chitov,chitov2}, but some traces of these phase transitions can still be found in the rung Berry phases of the ladder,
suggesting the presence of a series of 2S cross-overs in spin-S ladders. 

The paper is organized as follows. In Sec.~\ref{sec:berry_phase}, we define the Berry phase in a spin system and recall some basic facts established in Refs. \cite{hatsugai,hirano_katsura_hatsugai}. In Sec. \ref{sec:ffl} we show that the Berry phase can be used to establish the phase diagram of a frustrated ladder. We then turn in Sec. \ref{sec:ladder} to the investigation of the model which interpolates between non-frustrated spin-S ladders and dimerized spin-S chains. Some details about the Berry phase calculations are given in three appendices.

\section{Berry phases}
\label{sec:berry_phase}
The Berry phase can in principle be defined for any Hamiltonian $H(\phi)$ which depends periodically on a parameter $\phi$\cite{berry}.
If $| GS(\phi)\rangle $ denotes a single-valued ground state of  $H(\phi)$, the Berry connection is defined by
$A(\phi)=\langle GS(\phi)|\partial_{\phi}|GS(\phi)\rangle$, and the Berry phase is the integration of the Berry 
connection over a loop: $i\gamma=\oint A(\phi)d\phi$. A few years ago,
Hatsugai \cite{hatsugai} has shown that one can detect the presence of a singlet on a given bond of an antiferromagnet by calculating the Berry phase associated to a twist of the transverse component of the spin-spin interaction on this bond:
\begin{eqnarray}
\label{eq:twist}
S^+_iS^-_j+S^-_iS^+_j \rightarrow e^{i\phi}S^+_iS^-_j+e^{-i\phi}S^-_iS^+_j.
\end{eqnarray}
In a VBS state, the Berry phase is related to the number $N_s$ of singlets on the bond by: 
\begin{eqnarray}
 \gamma=N_s\ \pi ~~~~(\mathrm {mod} \ 2\*\pi).
\end{eqnarray} This quantization is protected by time-reversal symmetry. However, if the gap closes on the path, $|GS(\phi)\rangle$ is not well defined anymore and the Berry phase takes a random value. In this case, we say that the Berry phase is undefined. This happens in particular at quantum phase transitions since in that case the gap closes at $\phi=0$.

This criterion has been tested on dimerized spin chains with larger spins\cite{hirano_katsura_hatsugai}, which undergo a series of quantum phase transitions upon increasing the dimerization\cite{affleck_haldane}, and indeed the Berry phase jumps between $0$ and $\pi$ at each transition, in agreement with the interpretation of the phases in terms of valence-bond singlets.

\section{Fully frustrated spin-1 ladder with bilinear-biquadratic interaction.}
\label{sec:ffl}
\begin{figure}[t]
\includegraphics[width=0.48\textwidth]{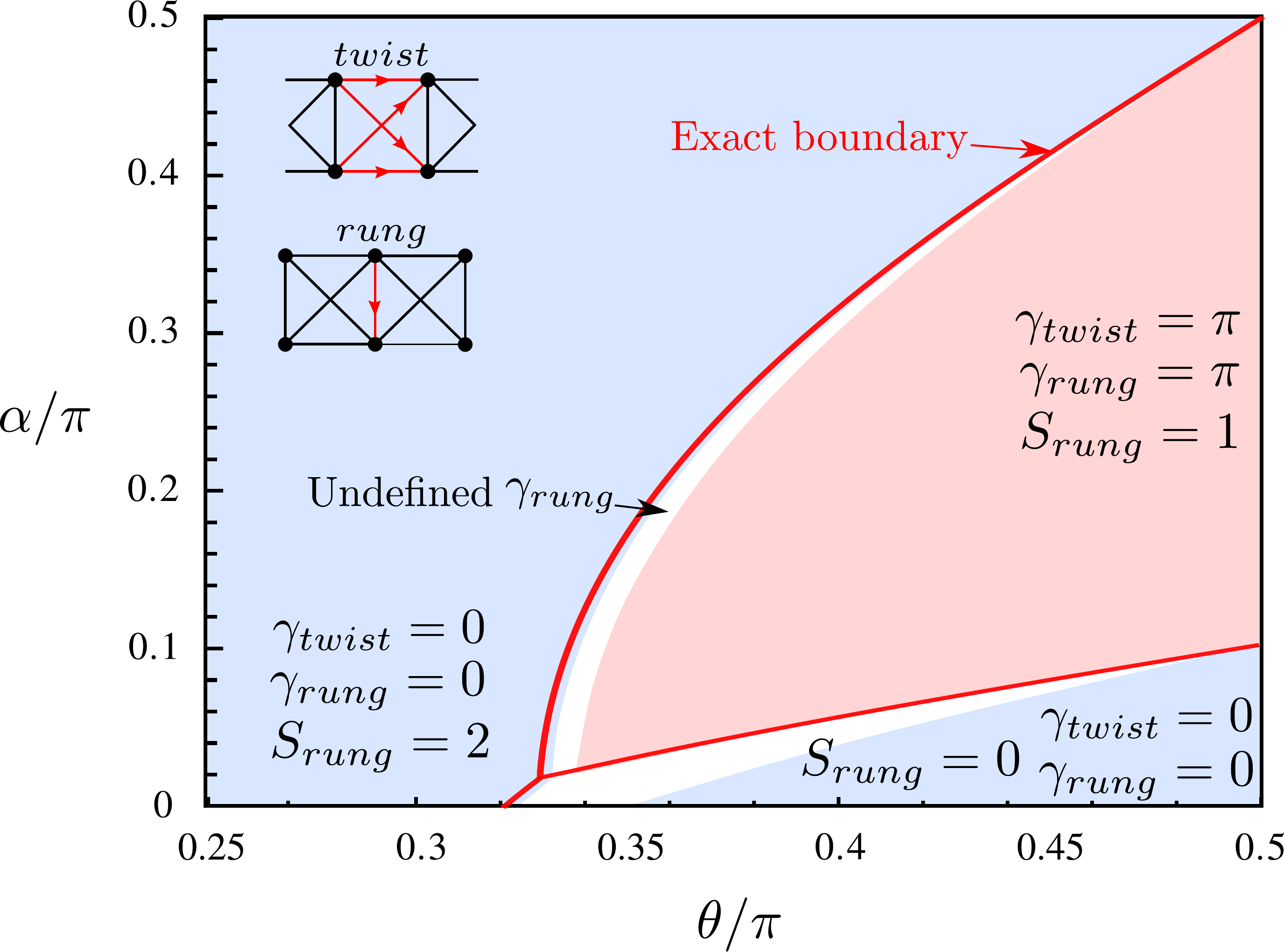}
\caption{(Color online) Phase diagram of the frustrated spin-1 two-leg ladder defined by the Hamiltonian (\ref{eq:B})
with the convention $J_\parallel=J\cos\theta$ and $J_\perp=J\sin\theta$. The bilinear coupling along the rung is equal 
to $J_\perp \cos\alpha$ and the biquadratic one to $J_\perp \sin\alpha$. Bold (red) lines are the exact phase boundaries. In the light blue regions, $\gamma_{rung}=\gamma_{twist}=0$, while in the light red region  $\gamma_{rung}=\gamma_{twist}=\pi$. In the white region, the rung Berry phase is undefined. For the twist Berry phase, the region where it is undefined is much smaller. Its extent is comparable to the thickness of the line which represents the exact boundaries (see Appendix C).}
\label{fig:FFSL}
\end{figure}

To get some insight into which Berry phases might be useful in the investigation of spin ladders, we start with a model whose phase diagram is known exactly, namely the fully frustrated spin-1 ladder with bilinear and biquadratic interactions on the rungs defined by the Hamiltonian:
\begin{eqnarray}\label{eq:B}
H&=&J_\parallel\sum_{i}\left({\bf S}_{i,1}+{\bf S}_{i,2}\right)\cdot\left({\bf S}_{i+1,1}+{\bf S}_{i+1,2}\right)\nonumber\\
&+&J_{\perp}\sum_i\left[\cos\alpha({\bf S}_{i,1}\cdot{\bf S}_{i,2})+\sin\alpha({\bf S}_{i,1}\cdot{\bf S}_{i,2})^2\right]
\end{eqnarray}
The total spin on each rung commutes with the Hamiltonian and is therefore a conserved quantity. 
The energy of this Hamiltonian is minimal when the total spin is the same on all rungs, and accordingly the ground state can be effectively described either as a product of singlets on the rung, or as a spin-1 chain, or as a spin-2 chain. For the singlet phase, the ground state energy per bond is given by $E_S=J_\perp(-2\cos\alpha+4\sin\alpha)$, for the triplet phase by $E_T=J_\perp(-\cos\alpha+\sin\alpha)+J_\parallel E_1$ and for the quintuplet phase by $E_Q=J_\perp(\cos\alpha+\sin\alpha)+J_\parallel E_2$, where $E_1\approx-1.401$ \cite{white_huse} and $E_2\approx-4.761$ \cite{PhysRevB.55.2721} are the ground state energies per site of the $S=1$ and $S=2$ Heisenberg chains in units of the coupling constant. Using these energies, we  determined the exact phase diagram as a function of $\theta$ and $\alpha$ (see Fig.\ref{fig:FFSL}). It consists of three phases with respectively total spin 0, 1 or 2 on every rung. Note that the intermediate phase
with spin-1 on each rung is not stabilized in the absence of a biquadratic interaction, so that we had to include one to be able to discuss this phase.

To test which Berry phases are most appropriate to detect such phase transitions, we have calculated various Berry phases corresponding to one or several simultaneous local twists, and two of them turned out to be relevant, namely the twist and rung Berry phases. The twist Berry phase is obtained by introducing the same local spin twist as described by Eq. \ref{eq:twist} on all couplings between a pair of rungs. The rung Berry phase is obtained by introducing a local spin twist on one rung, both in the bilinear and in the biquadratic coupling. A pictorial representation of these Berry phases is given in the insets of Fig.\ref{fig:FFSL}. A red arrow represents a local twist on one bond. 

As shown in Fig.\ref{fig:FFSL}, both Berry phases are equal  to $\pi$ in the spin-1 phase and to 0 in the spin-0 and 2 phases, up to some
region shown in white where they are undefined. These values of the Berry phases can be easily explained by the valence-bond singlet picture. First of all,
we expect both Berry phases to be zero in the singlet and quintuplet phase. Indeed, the singlet phase can be effectively represented by two spin-1/2 singlets on each rung and no singlet on the legs or diagonal bonds, while the quintuplet phase is adiabatically connected to a phase which contains singlets on each bond of the leg or on each diagonal bond and no singlet on the rung. Both lead to an even number of singlets on the rung and on the bonds between two rungs, leading to 0 rung and twist Berry phases. By contrast, the triplet phase contains one singlet on the rung, leading to a $\pi$ rung Berry phase, and one singlet on one of the legs or diagonal bonds, which gives rise to a twist Berry phase of $\pi$. 

Note however that the rung and twist Berry phases are not equivalent {\it a priori}. In fact, 
while the region where the rung Berry phase $\gamma_{rung}$ is undefined is rather large, the region where the twist Berry phase  $\gamma_{rung}$ is undefined is very small and cannot be seen at the scale of the graph. 
This means that the finite size effects for the twist Berry phase are much smaller than for the rung Berry phase which, as we shall see in the context of non-frustrated spin ladders,
seems to be a general property of the rung and twist Berry phases. So the fact that, up to the small regions where they are undefined,
the rung and twist Berry phases lead to the same phase diagram is not a trivial result, and as we shall see, it does not carry over
to non-frustrated ladders.

\section{From dimerized spin-chains to non-frustrated ladders}
\label{sec:ladder}
\subsection{Berry phases}
\label{sec:ladder_bp}
\begin{figure}[t]
\includegraphics[width=0.3\textwidth]{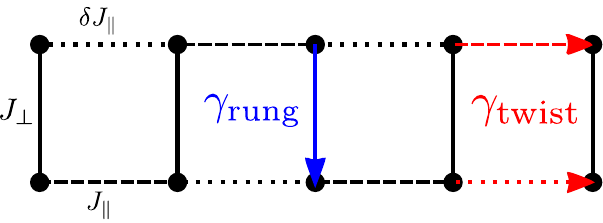}
\caption{(Color online) Sketch of the dimerized ladder that interpolates between the bond-alternating chain  ($\delta=0$) and the two-leg 
ladder ($\delta=1$). 
Blue and red arrows represent the ways to introduce the local twists corresponding to the rung and twist Berry phases.}
\label{fig:ladder_chain}
\end{figure}

Next, we turn to the investigation of a model which interpolates between a spin-S chain and a standard two-leg ladder (see Fig.\ref{fig:ladder_chain}). This system is defined by the following Hamiltonian:
\begin{equation}
H=J_{\perp}\sum_i{\bf S}_{i,1}\cdot{\bf S}_{i,2}+$$
$$\frac{J_\parallel}{2}\sum_i\sum_{\alpha=1,2}\left((1+\delta)+(-1)^{i+\alpha}(1-\delta)\right){\bf S}_{i,\alpha}\cdot{\bf S}_{i+1,\alpha}.
\end{equation}
In the limit $\delta=0$ the system is a dimerized chain and in the limit $\delta=1$ a non-frustrated ladder.

We have calculated the rung and the twist Berry phases of this model for S=1/2, 1, 3/2 and 2 for systems with up to 24, 12, 12 and 8 sites respectively. For the rung Berry phase and for $S=1/2$, we have also performed DMRG calculations up to 100 sites (see Appendix C). 
  The results are summarized in Fig.~\ref{fig:Diagrams}. In these phase diagrams, the regions where the twist Berry phase
is equal to $\pi$ are shown in light red, while those where the rung Berry phase is equal to $\pi$ are shown in light blue. Remarkably
enough, the phases defined by rung and twist Berry phases are no longer equivalent, except close to the limit of dimerized chains. 
That the phases have to be the same in the case $\delta=0$, i.e. for dimerized chains, is clear. Indeed, the rung and twist Berry phases are local phases on neighboring bonds on a chain in that case, and, according to the valence-bond picture, they carry complementary information and must lead to the same transitions. Within the precision of our numerical results, this remains true only up to some critical value beyond which the two boundaries separate. This critical value seems to decrease for higher spin. It is given for small spin by $\delta_c\approx0.4$ for $S=1/2$ and $\delta_c\approx0.2$ for $S=1$. For $\delta<\delta_c$, the system is apparently topologically equivalent to the dimerized chain. Beyond that value, the boundaries become progressively very different, and for the standard ladder at $\delta=1$,
the difference becomes qualitative: the rung Berry phase still undergoes the same number of transitions as for the dimerized chain,
while the twist Berry phase does not undergo any transition in the range $\theta \in (0,\pi/2]$.

\begin{figure}[h!]
\includegraphics[width=0.5\textwidth]{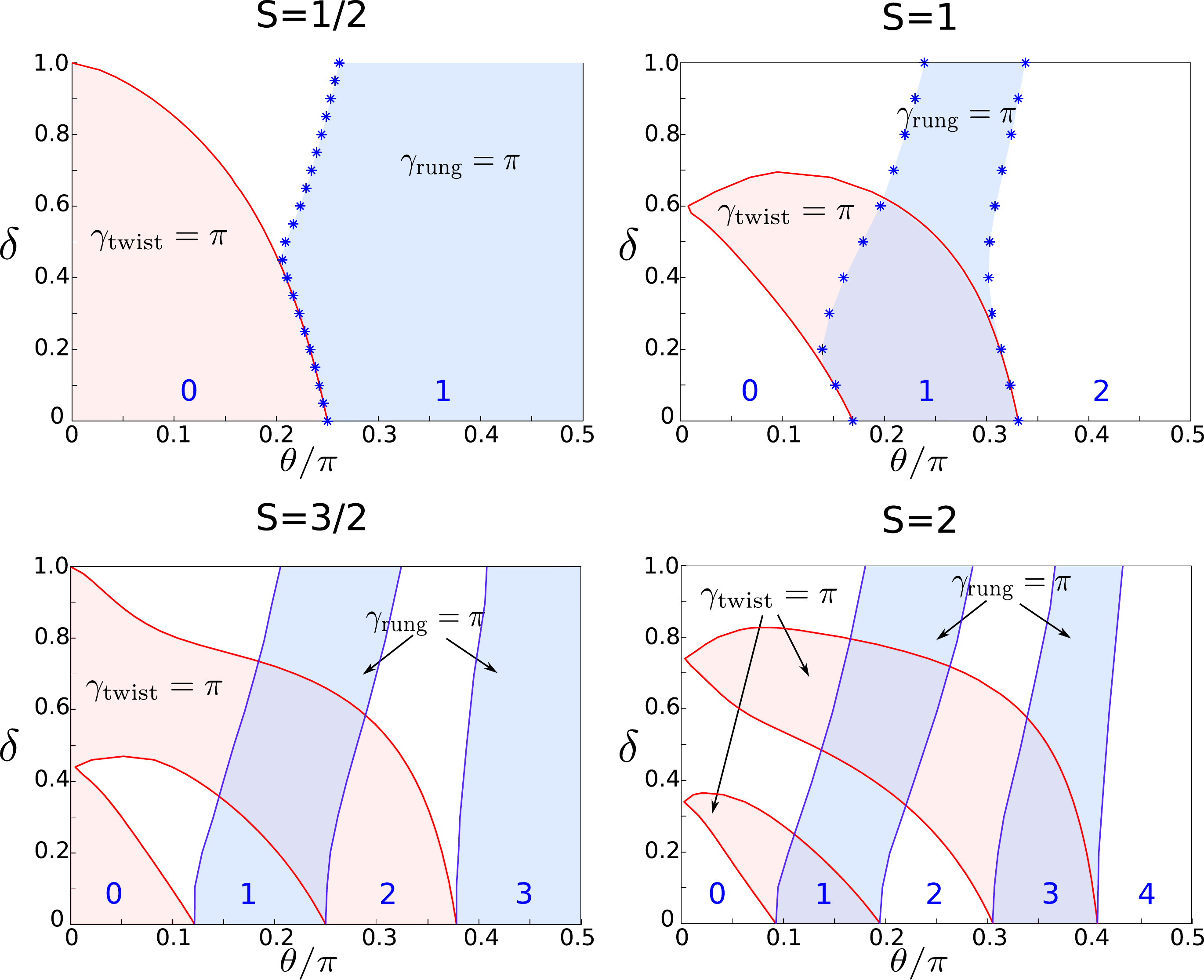}
\caption{(Color online) Phase diagrams of the spin-S dimerized ladder for $S=1/2$, $1$, $3/2$ and $2$ as a function of $\theta$ with the convention $J_\perp=J \sin \theta$ and $J_\parallel=J\cos \theta$. The parameter $\delta$ varies from 0 (dimerized chain) to 1 (2-leg ladder). The rung Berry
phase is equal to $\pi$ in the light (blue) regions and to $0$ elsewhere, while the twist Berry phase is equal to $\pi$ in the dark (red) regions and to $0$ elsewhere. The integers 0, 1... correspond to the number of valence-bond singlets on the rungs according to the valence-bond interpretation of the dimerized chains. The phase boundaries are indicated by dots when they were obtained by a finite-size analysis, and by colored lines when they were obtained with a single system size. The number of sites of the largest system reached for $S=1/2$ is $N=24$ for twist Berry phase and $N=100$ for the rung Berry phase, while for $S=1$, $S=3/2$ and $S=2$ the largest number of sites is $N=12$, $N=12$ and $N=8$ respectively for both rung and twist Berry phases.}
\label{fig:Diagrams}
\end{figure}
Before embarking on the physical interpretation of these results, let us make a few additional comments. First of all, 
a ladder with ferromagnetic rung coupling is equivalent to a $2S$ Haldane chain. The Berry phase of such a chain is $0$ if $2S$ is even and $\pi$ if $2S$ is odd. This region corresponds in our case to $\theta<0$. For half-integer spins, the twist Berry phase is equal to $\pi$ along the line $\theta=0$, except at $(2S+1)/2$ points where it is undefined. For integer spins, it is zero on the line $\theta=0$, except at $S$ critical points where it is again undefined. 

\begin{figure}[t]
\includegraphics[width=0.4\textwidth]{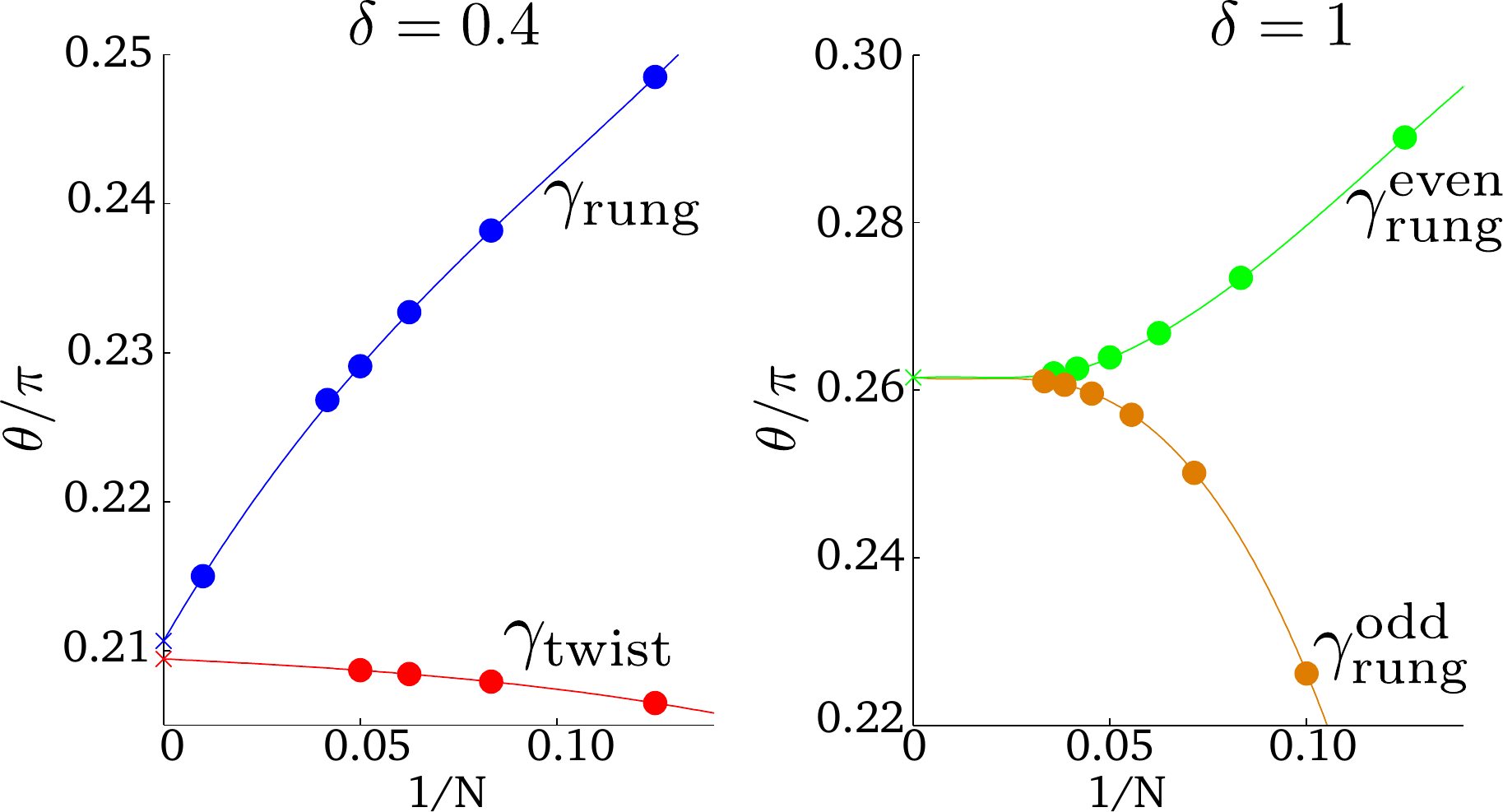}
\caption{(Color online) Examples of finite-size scaling of the Berry phase transitions of the spin-1/2 ladder as a function of the inverse number of sites. Left panel: Critical $\theta$ for the twist and rung Berry phases in the ladder with dimerization parameter $\delta=0.4$. Right panel: Critical $\theta$ for the rung Berry phase in the standard ladder $\delta=1$ with even and odd numbers of rungs. 
The results have been fitted with a polynomial in $1/N$.}
\label{fig:Scaling}
\end{figure}

Besides, as already mentioned in the case of the fully frustrated ladder, the twist Berry phase has smaller finite size effects than the rung Berry phase. As an example, we show the finite-size scaling for twist and rung Berry phases at $\delta=0.4$ on Fig.~\ref{fig:Scaling}. For standard ladders ($\delta=1$), the Hamiltonian becomes invariant under a translation of two sites, i.e. of one rung, which is not the case for $\delta<1$. The unit cell therefore contains only two sites and not four, and the number of unit cells can be both even or odd.  There is a strong even-odd effect, but taking into account only odd or only even system sizes leads to the same critical point in the thermodynamic limit.

\subsection{Twist Berry phase, gap closing and edge states}

\begin{figure}[t]
\includegraphics[width=0.5\textwidth]{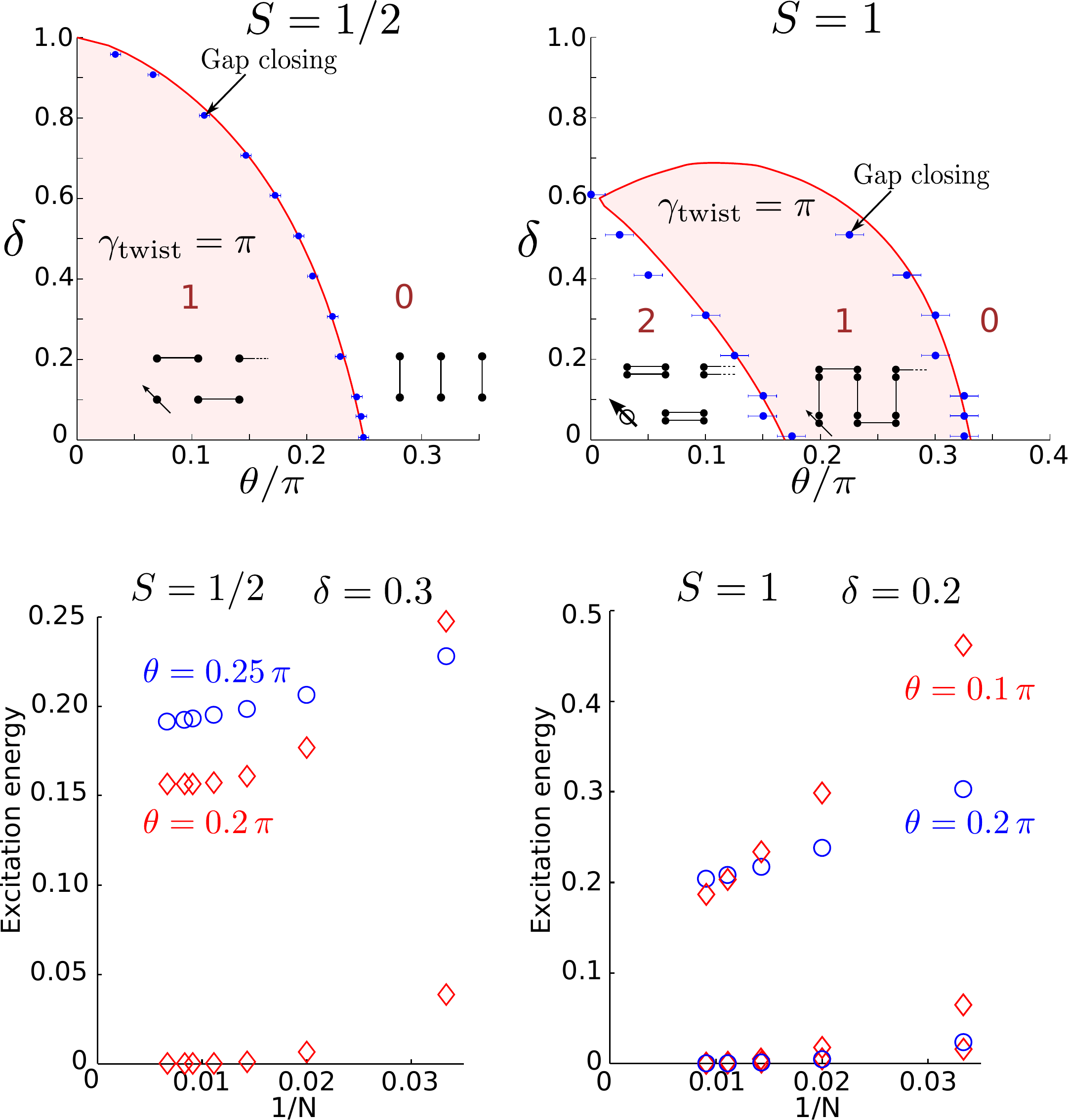}
\caption{(Color online) Upper panel: Phase diagrams of the spin $1/2$ and spin $1$ dimerized ladders. The twist Berry phase is equal to $\pi$ in the
dark (red) region and to $0$ elsewhere. The blue circles denote points where the system is gapless. For S=1/2 the critical values of $\theta$ are the result of a finite-size analysis of systems with up to $150$ sites, while for S=1 they correspond to a system with $90$ sites. The integers in each phase indicate the number of low-lying excited states in the subspace $S_{tot}^z=0$ for a dimerized ladder with open boundary conditions. Lower panel: Low-energy spectrum of the spin $S=1/2$ dimerized ladder at $\theta=0.2\,\pi$ and $\delta=0.3$ (left) and of the $S=1$ dimerized ladder at $\theta=0.2\,\pi$ and $\delta=0.1$ (right) in the subspace $S_{tot}^z=0$}
\label{fig:Hald+edge}
\end{figure}

For dimerized chains, the phase transitions at which the Berry phases change are known to be quantum phase
transitions with a gap closing. As a first step towards the interpretation of the results of the previous section, we
have calculated with DMRG the gap in the $(\theta,\delta)$ plane for $S=1/2$ and $S=1$, and we have mapped
out the phase boundaries at which the gap closes. 
For $S=1/2$ the points where the bulk gap closes were obtained from a finite-size scaling analysis of systems with up to $150$ sites. For the $S=1$ case we show the results for ladders with up to $90$ sites. We kept up to $1000$ states, and with this number of states the energy and the gap were well converged. As seen in Fig.~\ref{fig:Hald+edge}, the boundaries defined by the gap closing are consistent with those
defined by the twist Berry phase, and {\it not} with those defined by the rung Berry phase. This is particularly clear for spin 1/2, 
where DMRG leads to a very precise boundary for the gap closing, but this is also quite clear for spin-1 for the left boundary, and we
expect this to be true for larger spin as well. 

From these results, we conclude that changes in the twist Berry phase signal quantum phase transitions at which the gap closes. The big advantage of the twist Berry phase in investigating such phase transitions is that the results are already very accurate for small systems. The twist Berry phase method is in fact related to a level crossing analysis with twisted boundary conditions, a method 
called level spectroscopy and known to give accurate results already for small system sizes\cite{okamoto,nomura}. 

Edge states are another characteristics of valence-bond solids. In the spin-1 chain, it has indeed long been known that edge states appear in finite chains \cite{kennedy,hagiwara,ng,lecheminant_orignac} with open boundary conditions, effective spin-1/2 degrees of freedom
located at the ends of the chains building a singlet and a triplet - the Kennedy triplet \cite{kennedy} - whose energy become degenerate in the thermodynamic limit. 

In the present case, edge states are expected to appear if we consider systems with open boundary conditions and vertical edges,
and again extensive numerical simulations for spin 1/2 and spin 1 have shown that the boundaries defined by changes in the edge-state structure correspond to those where the twist Berry phase changes between $0$ and $\pi$.
For spin $S=1/2$, the region where there are low-lying triplet excitations coincides with the region where the twist Berry phase is $\gamma_{twist}=\pi$. Edge states disappear at the line where the twist Berry phase changes from $\pi$ to $0$. This is illustrated in the bottom left panel of Fig.~\ref{fig:Hald+edge}, where we show the gap of low lying excited states in the sector $S_{tot}^z=0$ for different system sizes for two different points in parameter space. Before the transition, at $\delta = 0.3$, $\theta=0.2\pi$, we have one edge state. After the transition, at $\delta = 0.3$, $\theta=0.25\pi$, we have no edge state anymore.

In the spin-1 case, the valence-bond picture predicts that, in the limit of small rung coupling and small $\delta$, there are effective spins 1 at the boundaries of the chain, leading to one low-lying quintuplet and one low-lying triplet excitation. 
Increasing the coupling on the rungs leads to a different valence-bond structure that leaves effective spins 1/2 at the end, leading to a 
low-lying triplet excitation. Increasing further the rung coupling, singlets appear on the rung, and all edge states disappear.
These results are confirmed by the case shown in the bottom right panel of Fig. \ref{fig:Hald+edge}: we have two edge states before the transition, at $\delta = 0.2$, $\theta=0.1\pi$, and only a single edge state after the first transition at  $\delta = 0.2$, $\theta=0.2\pi$.

\subsection{Rung Berry phase, cross-overs and impurity healing}

Since the rung and twist Berry phases define different regions in parameter space, and since the twist Berry phase keeps track
of the quantum phase transitions at which the gap closes, the next step is to investigate the physical implications (if any) of the change of rung Berry phase.  In the standard ladder, we observe no change in the twist Berry phase when increasing $\theta$, except at the gapless 
point $\theta=0$ for half-integer spins (see Fig. \ref{fig:Diagrams}). This has to be contrasted to the rung Berry phase which change of value $2S$ times in the interval $(0,\pi/2]$. So the transitions detected by the rung Berry phase cannot be associated to real quantum phase transitions. Note that the transition at $\theta=0.26\,\pi$ for spin-1/2 ladder has already been reported in Ref.~\onlinecite{maruyama_hirano_hatsugai} in the context of an investigation of the frustrated spin-1/2 ladder with four-spin ring exchange. Note also that this transition does not show up in the string order parameters that distinguish the phases with ferromagnetic and antiferromagnetic rung couplings \cite{nishiyama,kim_fath_solyom_scalapino,fath_legeza_solyom}.

\begin{figure}[t]
\includegraphics[width=0.45\textwidth]{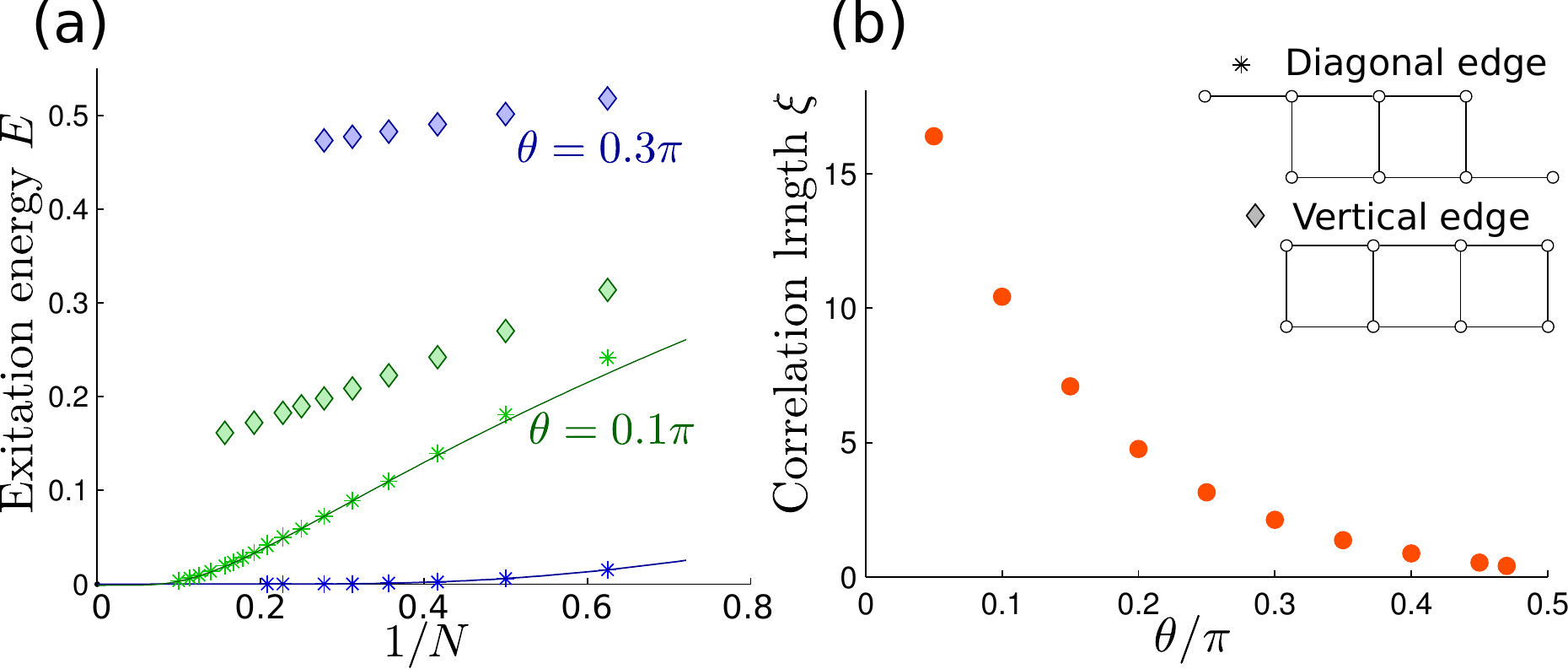}
\caption{(Color online) (a) Low energy spectrum of the spin-1/2 ladder with vertical edges (diamond symbols) and diagonal edges (star symbols) for two values of the rung coupling: $\theta=0.1\,\pi$ (green) and $\theta=0.3\,\pi$ (blue). (b) Correlation length as a function of $\theta$ deduced from the exponential scaling of the edge state gap for a ladder with diagonal edges. Inset: sketch of clusters with vertical and diagonal edges.}
\label{fig:EdgeStates}
\end{figure}

To further confirm the absence of phase transition, we have investigated the appearance of edge states in the ladder with open boundary conditions. We have considered two different types of open boundary conditions: vertical and diagonal edges (see Fig.\ref{fig:EdgeStates}). 
In the strong rungs limit, we expect to see low-lying excitations in the ladder with diagonal edges, since spins at the boundaries are isolated, while the boundary spins in the ladder with vertical edges are coupled and do not form a Kennedy triplet. What we observe by going to sufficiently large system sizes is that these edge states remain present down to $\theta=0$. As shown in Fig.~\ref{fig:EdgeStates} for both weak and strong rung coupling, the system has excitations which decay exponentially fast with the system size. These results are in agreement with the absence of a real phase transition. 
Fitting the low-lying excitation spectra with an exponentially decaying function, we extracted the correlation length as a function of the rung coupling.
Usually, at a gapless phase transition, the length scale diverges, however in our case there is no divergence except at the critical gapless point $\theta=0$. This means that the Berry phase change at $\theta=0.26\,\pi$ signals at  best a cross-over, but not a real quantum phase transition, as already noted in another context\cite{hatsugai_BEC_BCS}.

\begin{figure}[t]
\includegraphics[width=0.5\textwidth]{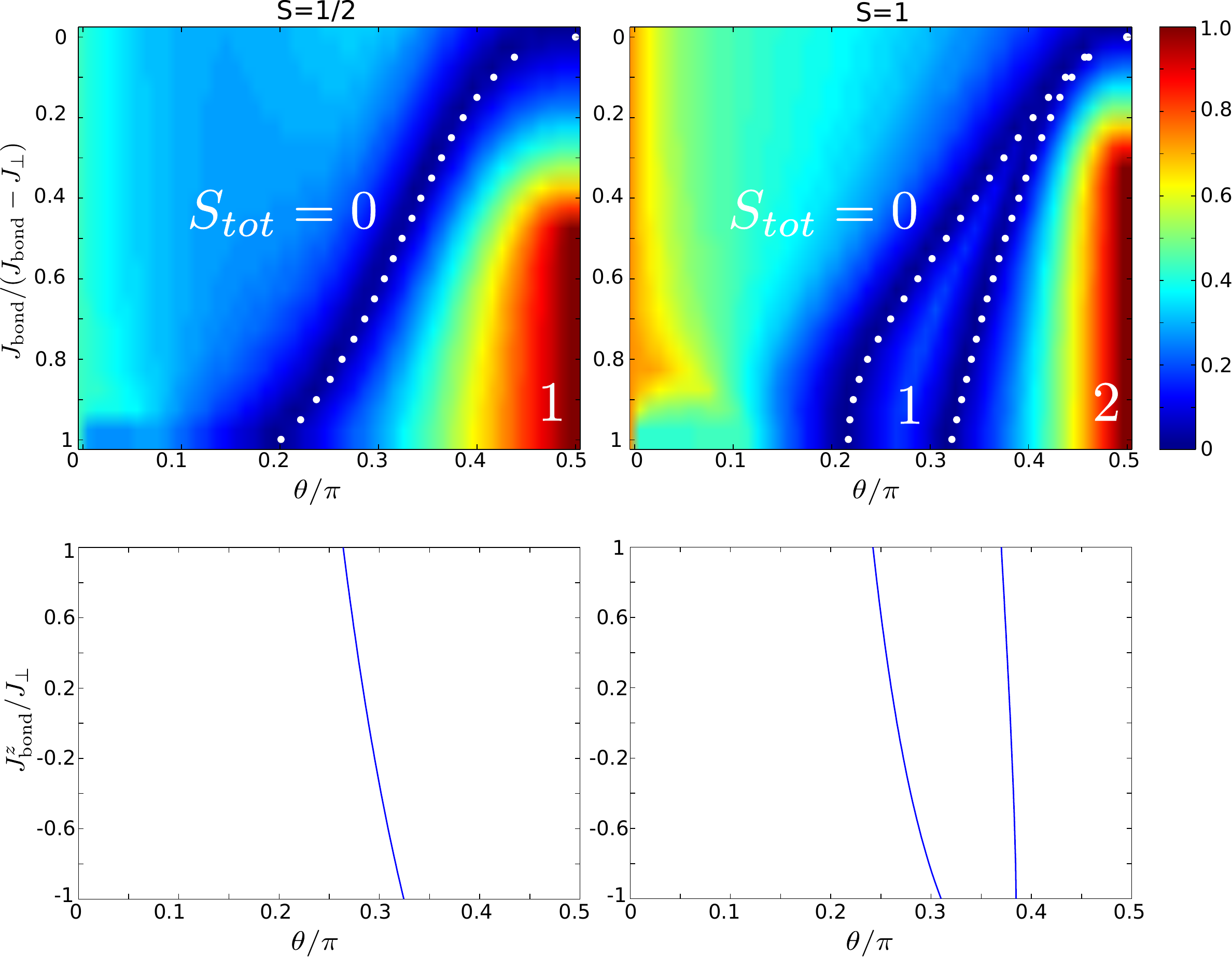}
\caption{(Color online) Upper panels: Energy gap between the ground state and the first excited state for regular antiferromagnetic spin-1/2 and spin-1 ladders with one ferromagnetic rung $J_{bond}$ with the convention $J_\perp=J \sin \theta$ and $J_\parallel=J\cos \theta$.  The white circles show the critical 
lines along which the system is gapless. The limit $J_{\mathrm{bond}}/(J_{\mathrm{bond}}-J_\perp)=1$ corresponds to replacing one rung by a spin-2S impurity. The colorbar on the right shows the values of the energy gap in units of $J$. Lower panels: Interpolation between a conventional twist of the transverse component of the spin-spin interaction on one rung ($J_{\mathrm{bond}}^z/J_\perp=1$) and a ferromagnetic rung  ($J_{\mathrm{bond}}^z/J_\perp=-1$). The blue lines are the critical lines where the systems are gapless. The  boundary conditions are periodic. The results were obtained with 20 sites for $S=1/2$ and 8 sites for $S=1$.}
\label{fig:Impurity}
\end{figure}

Coming back to the general spin-S case, the fact that the rung Berry phase undergoes $2S$ changes at values that evolve very smoothly
from the dimerized chain limit suggest that these changes correspond to cross-overs between regions with different values of the total spin of the rungs, as in the case of the spin-1 fully frustrated ladders, or equivalently with different numbers of valence-bond singlets on the rungs.
To confirm this picture, it would be nice to find a way to turn these cross-overs into real phase transitions. 

To achieve this, we first note that the change of rung Berry phase corresponds to a level crossing, hence to a real phase transition,
for a system with a twist $\phi = \pi$ imposed on one rung (see Appendix A for a discussion of the location of the singularities
that give rise to changes of the rung and twist Berry phases). This is not fully satisfactory because imposing a twist $\phi = \pi$
on a bond is equivalent to changing the sign of the coupling of the $x$ and $y$ components of the spin-spin interaction $J_{\mathrm{bond}}^x$ and $J_{\mathrm{bond}}^y$ while leaving that of the $z$ component $J_{\mathrm{bond}}^z$ unchanged, and this is not physically relevant, at least for quantum magnets where the spin-spin interaction is, to first approximation, isotropic in spin space. It would be physically more relevant to change the sign of $J_{\mathrm{bond}}^z$ as well.  
 
We were therefore led to study the ladder with one ferromagnetic bond $J_{\mathrm{bond}}<0$. 
The energy gap between the ground state and the first excited states for $S=1/2$ and $S=1$ are presented in the Fig.~\ref{fig:Impurity}. 
Quite remarkably, we have found that for any $J_{bond}<0$ there exist $2S$ gapless lines, which indicate transitions between states with different total spin, ranging from 0 for weak rung coupling to 2S for strong rung coupling. So, a spin-S ladder indeed undergoes a series of $2S$ phase transitions as a function of the rung
coupling provided one rung is kept fixed and ferromagnetic. The critical values obtained for 
$-J_{\mathrm{bond}}^{x,y}=J_{\mathrm{bond}}^z=J_\perp$, $J_{\mathrm{bond}}=-J_\perp$ and $J_{\mathrm{bond}}=-\infty$ 
are summarized in Table I. 

To check whether these phase transitions are related to the change of Berry phase, we have followed the level crossing observed
with a twist $\phi = \pi$ when changing the $z$ component of the spin-spin interaction from 1 to -1 (see Fig.~\ref{fig:Impurity}), and indeed
there is smooth evolution between them.

In the limit of a very strongly ferromagnetic bond, the system becomes equivalent to a 2S impurity in a spin-S ladder. This kind
of problem has been studied in the context of spin chains, and the ability for a system to screen an impurity and behave as a 
system without an impurity has been named healing\cite{eggert}. In the present case, the interpretation is that the heeling ability of a spin
ladder changes from very weak in the strong rung limit, with a ground state with total spin 2S, to very strong in the weak rung
limit, where the impurity is totally screened and the total spin is equal to 0. This ability is in turn a consequence of the effective
total spin of the rungs. If rungs are strong singlets, they are unable to couple to a magnetic impurity, but as they become more and 
more magnetic, they can couple to the impurity and finally completely screen it.

So, it appears that the changes of rung Berry phase are not associated to quantum phase transitions of the bulk system,
but that they signal changes in the nature of the local wave function of the rungs. These changes are progressive and are cross-overs rather than phase transitions, but they alter significantly enough the healing ability of the system to turn these cross-overs into phase
transitions in the presence of magnetic impurities.

\begin {table}[h!]

\begin{center}
\begin{tabular}{|c|c|c|c|}
  \hline 
 & Berry phase & FM rung & $2S$ impurity \\
 & $J_{\mathrm{bond}}^{x,y}=-J_\perp$ & $J_{\mathrm{bond}}=-J_\perp$ & $J_{\mathrm{bond}}=-\infty$ \\
 & $J_{\mathrm{bond}}^z=J_\perp$ &  &  \\
  \hline 
$S=1/2$ & $0.264\,\pi$ &  $0.324\,\pi$ &  $0.235\,\pi$\\
  \hline 
$S=1$ & $0.242\,\pi$ &  $0.310\,\pi$ &  $0.22\,\pi$\\
 & $0.370\,\pi$ &  $0.385\,\pi$ &  $0.32\,\pi$\\
  \hline 
\end{tabular}
\caption {Critical values of $\theta=\arctan (J_\perp/J_\parallel)$ for $S=1/2$ and $S=1$ deduced from the rung Berry phase, or associated to the introduction of a
ferromagnetic rung or of a spin-2S impurity.}
\end{center}
\end {table}

\section{Conclusion}

In the light of the present results, the Berry phase introduced by Hatusgai appears as a versatile and subtle tool to investigate 
quantum magnets. As we have seen, and as already noticed in another context\cite{hatsugai_BEC_BCS}, a change of Berry phase does not necessarily
imply the presence of a phase transition despite the fact that it is quantized. When it does signal a phase transition, as for instance 
the twist Berry phase for spin ladders, then it is a very efficient tool: finite-size effects are small so that accurate results can 
already be obtained for small systems, and the physical interpretation provided by the valence-bond singlet picture is quite transparent.
When it does {\it not} signal a phase transition, then it can reflect subtle aspects of the local physics of the system, still in connection with
the effective number of singlets on some bonds. 

In the case of spin-S ladders generalized to make a connection with dimerized spin chains, previously known results for the spin-1/2 and 
spin-1 cases have been recovered by investigating the Berry phase associated to a twist of all bonds between two rungs, and the simplicity
of the method as compared to other characterizations of these phase transitions has allowed us to generalize these results to $S=3/2$ and $S=2$. 

A rather different, and to a certain extent complementary, information has been extracted from the Berry phase associated to the twist of one rung of the ladder. In that case, the changes of Berry phase are not associated to phase transitions of the bulk system, but they reflect the
effective local nature of the rungs, with implications for the response of the system to a local magnetic impurity.

Whether or not a change of Berry phase signals a true phase transition is of course a very important issue. In view of the differences between the twist and rung Berry phases regarding the nature and location of the singularities that lead to a change of Berry
phase (see Appendix A), it is tempting to speculate that it may be possible in general to relate the nature of the singularities with the occurrence of a quantum 
phase transition. This is left for further investigation however.

\subsection{Acknowledgments}
We acknowledge useful discussions with Edmond Orignac and Karlo Penc.
We are also grateful to Salvatore Manmana
for letting us use his DMRG code for some of the DMRG calculations reported in this paper. The other DMRG calculations have been done with the ALPS library\cite{Alps,Alps_Bauer}.
This work has been supported by the Swiss National Fund.

\vskip1.cm
\begin{center}
{\bf APPENDIX A: Berry phase and singularities }
\end{center}
To discuss the origin of the changes of the various Berry phases, it is convenient to introduce a generalized modification of the 
$x$ and $y$ couplings on a bond according to $S^+_iS^-_j+S^-_iS^+_j \rightarrow K S^+_iS^-_j+K^{*}S^-_iS^+_j$, 
where $K$ is an arbitrary complex number and not just a phase factor. With this definition, the Berry phase is 
related to the integral of the Berry connection along the unit circle in the complex plane of $K$, but one can calculate
the Berry phase associated to any contour in that plane. Then, according to a general result for planar contours\cite{berry}, 
the Berry phase will be equal to $0$ if the gap does not close inside or on the contour, it will be equal to $\pi$
if the gap closes at one point inside the contour, and more generally it will be equal to $n\pi\  (\mathrm{mod}\ 2\pi)$ if the
gap closes at $n$ points inside the contour.  Since the Berry connection cannot be defined
at a point where the energy gap closes, we call such a point a singularity. Then the Berry phases studied in the present
paper are completely defined by the number of such singularities inside the unit circle.

\begin{figure}[t]
\includegraphics[width=0.45\textwidth]{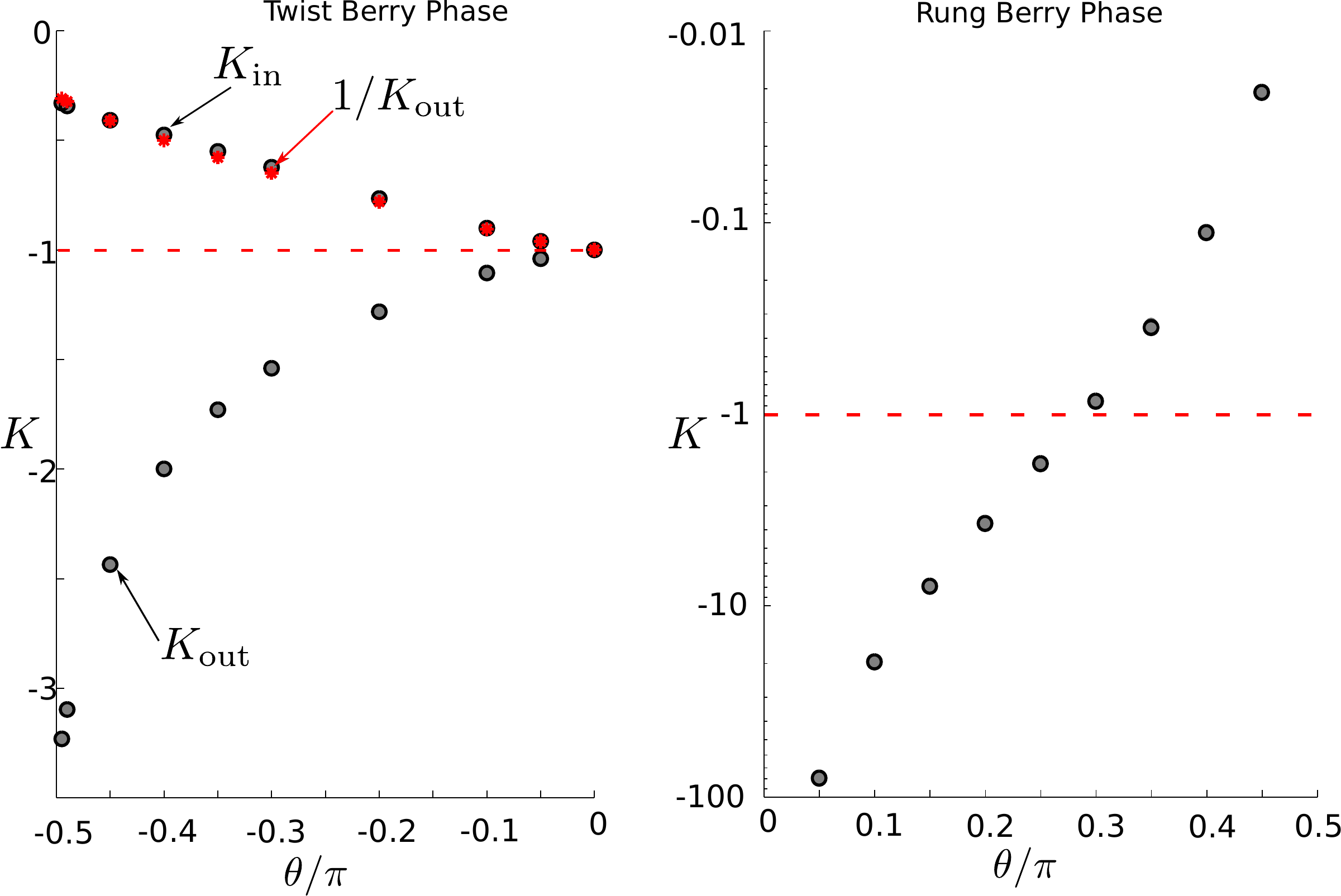}
\caption{(Color online) Positions of the level crossings for a S=1/2 ladder as a function of $\theta$ with the convention $J_\perp=J \sin \theta$ and $J_\parallel=J\cos \theta$.
A purely real modification $K$ of the coupling constant of the $x$ and $y$ components is applied on the two bonds corresponding to the 
twist Berry phase (left panel) and on a rung (right panel).  The dashed (red) line corresponds to the $\phi=\pi$ point on the integration contour used for the calculation of the Berry phase. The positions of the singularities on the real axis are represented as grey circles. The red stars on the left panel corresponds to the inverse of the positions of the singularity outside the unit circle $K_{\mathrm {out}}$ and coincide within numerical accuracy with the critical points inside it $K_{\mathrm{in}}$}.
\label{fig:Singularity}
\end{figure}

For unfrustrated ladders, and more generally for the model that interpolates between the dimerized chain and the ladder, 
we found that the singularities are always located on the real axis, i.e. for $K$ real. However, there is a significant
difference between the twist and the rung Berry phases, as shown in Fig.~\ref{fig:Singularity}. For the twist Berry phase, there
is no singularity for $\theta>0$, a singularity appears at  $K=-1$ for $\theta=0$, and it then splits into two singularities, 
one at $K_{\mathrm{out}}<-1$, and one at $-1<K_{\mathrm{in}}<0$ such that $K_{\mathrm{in}}\approx1/K_{\mathrm{out}}$. By contrast,
there is always a singularity for the rung Berry phase in the parameter range $\theta>0$, and, coming from very negative
values at small $\theta$, it simply crosses the point $K=-1$ at $\theta=0.26\,\pi$. There is thus a qualitative difference in that case in the
way a singularity appears inside the unit circle between the twist Berry phase, whose change signals a true phase transition, and the 
rung Berry phase, whose change can only be associated to a cross-over. It would be interesting to see if this observation
can be generalized to other situations.

\begin{figure}[t]
\includegraphics[width=0.45\textwidth]{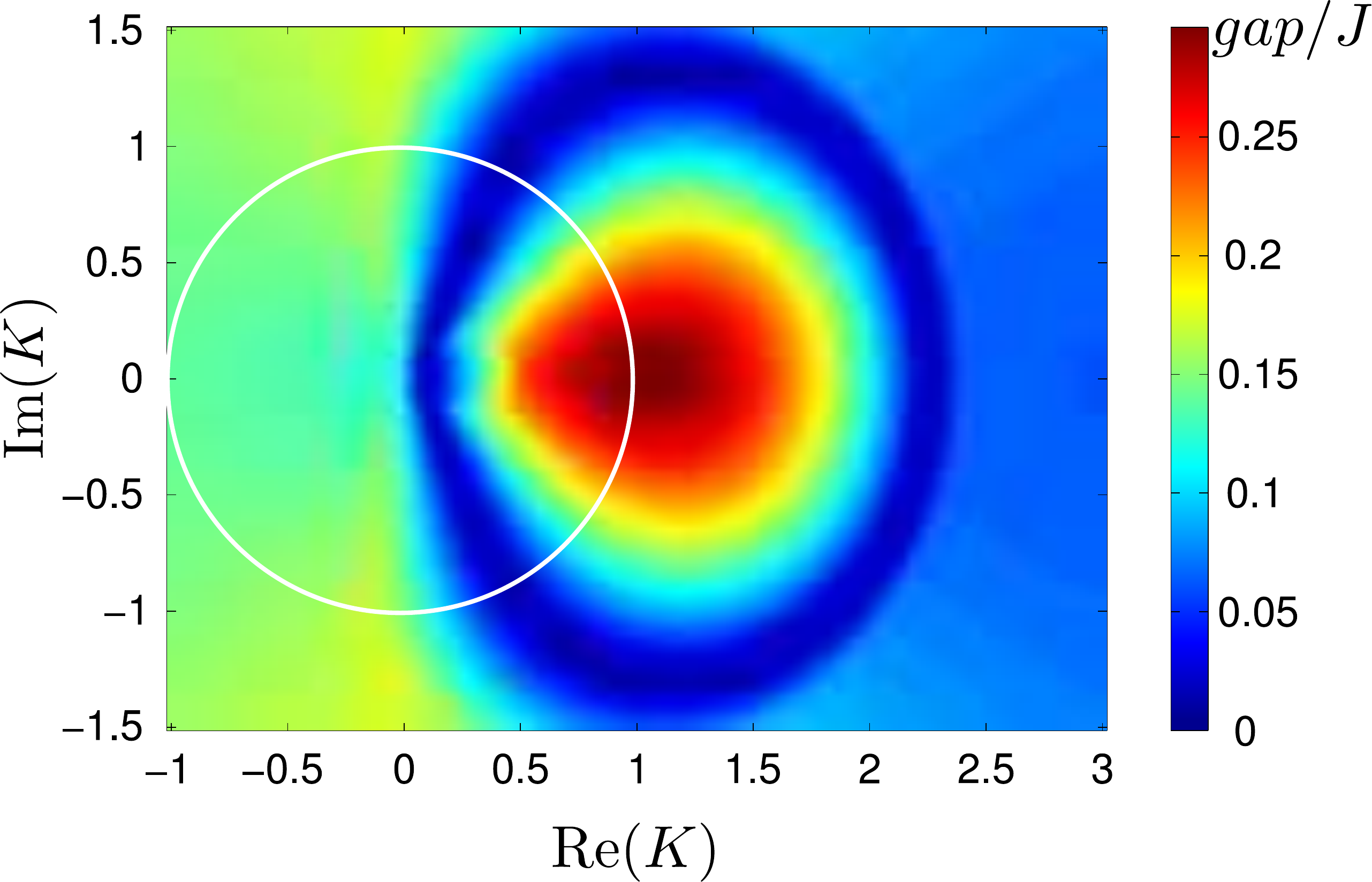}
\caption{ (Color online) Energy gap between the ground state and the first excited state for the fully frustrated spin-1 ladder described by Eq. \ref{eq:B} for $\alpha = 0.05 \, \pi$ and $\theta = 0.44\,\pi$ with generalized parameter $K$ applied on the rung. The white circle indicates the integration path used to compute the Berry phase.} 
\label{fig:FFSLSingularity}
\end{figure}

\vskip1.cm
\begin{center}
{\bf APPENDIX B: Finite-size effects for the fully frustrated spin-1 ladder}
\end{center}

In the case of the frustrated spin ladder, both Berry phases turn out to be undefined over a finite portion of the phase diagram and not at a single line. This is related to the fact that, instead of isolated singularities, there are gapless lines in the complex parameter space (see Fig. \ref{fig:FFSLSingularity}). As long as such a line is inside or crosses the unit-circle, the Berry phase is undefined.

\begin{figure}[t]
\includegraphics[width=0.45\textwidth]{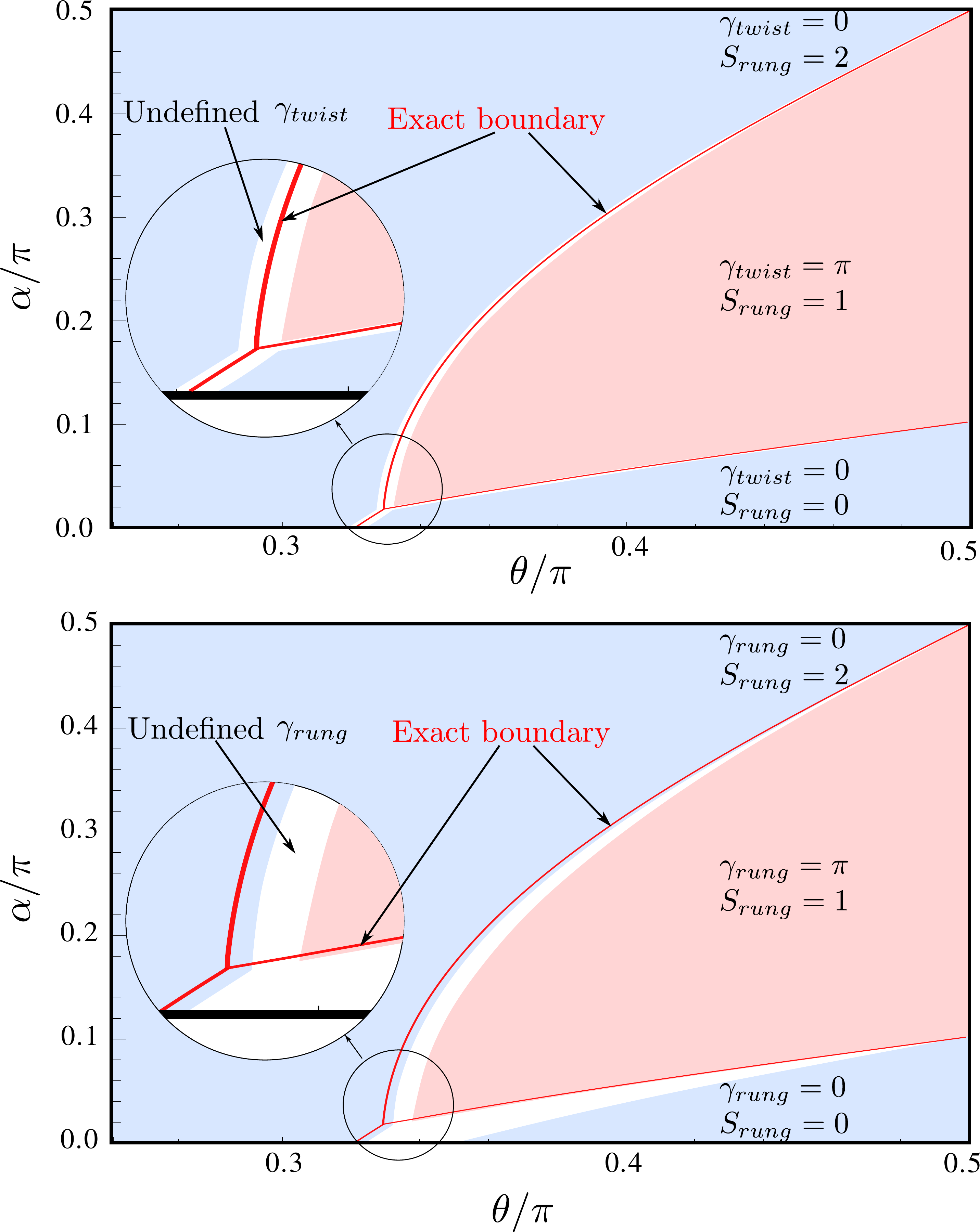}
\caption{(Color online) Phase diagram of the frustrated spin-1 two-leg ladder defined by the Hamiltonian (\ref{eq:B})
with the convention $J_\parallel=J\cos\theta$ and $J_\perp=J\sin\theta$. The bilinear coupling along the rung is equal 
to $J_\perp \cos\alpha$ and the biquadratic one to $J_\perp \sin\alpha$. Bold (red) lines are the exact phase boundaries. In the light blue regions, $\gamma=0$, while in the light red regions  $\gamma=\pi$. In the white region, the Berry phase is undefined. Top panel: twist Berry phase; bottom panel: rung Berry phase.}
\label{fig:ffl_zoom}
\end{figure}

In Fig.~\ref{fig:ffl_zoom}, we show enlarged versions of the phase diagrams obtained with the two Berry phases. Both are undefined in
regions close to the boundaries of the true phase transition. Note however that the twist Berry phase is undefined in a much smaller region than the rung Berry phase. Moreover, the true boundaries is always included in the region where the twist Berry phase is undefined, while for the rung Berry phase the exact boundaries are sometimes outside this region. This is very probably a finite size effect.

\vskip1.cm
\begin{center}
{\bf APPENDIX C: Some technical details about the numerical determination of the Berry phase}
\end{center}

The numerical calculation of the Berry phase has been done following the prescription of Ref.~\cite{hirano_katsura_hatsugai}. To get well
converged results, it turned out to be sufficient to discretize the integral as a sum of $16$ terms. However, since one needs the 
Berry phase for
several parameters, this becomes quite expensive numerically for large system sizes because, altogether, one has to diagonalize the Hamiltonian a significant number of times. Now, as we saw, a change of Berry phase is always associated to a level crossing along the integration path, and this level crossing usually appears for simple values of the twist parameter \cite{hirano_katsura_hatsugai_gap}, $\phi=0$ or $\phi = \pi$ . It is therefore sufficient to compute the energy of the ground state and of the first excited state of the system for a single value of $\phi$, either $\phi=0$ or $\phi = \pi$ depending on the system, and to determine the parameter at which a level crossing occurs, thus saving a lot of unnecessary calculations. Moreover, this method allows to compute the transition point for larger systems with DMRG. Computing the Berry phase with DMRG would be difficult because there is no direct representation of the ground state. However, the level crossing for $\phi=\pi$ can be computed for much larger system sizes than the one accessible with exact diagonalizations. This method was used for example to establish precisely the boundary of the $\gamma_{rung}=\pi$ phase for spin $S=1/2$ in Fig.~\ref{fig:Diagrams}.

\bibliographystyle{prsty}
\bibliography{bibliography}

\begin{thebibliography}{10}

\bibitem{thouless}
D.~J. Thouless, M. Kohmoto, M.~P. Nightingale, and M. den Nijs, Phys. Rev.
  Lett. {\bf 49},  405  (1982).

\bibitem{avron}
J.~E. Avron, D. Osadchy, and R. Seiler, Physics Today {\bf 56},  38  (2003).

\bibitem{hasan_kane}
M.~Z. Hasan and C.~L. Kane, Rev. Mod. Phys. {\bf 82},  3045  (2010).

\bibitem{qi_zhang}
X.-L. Qi and S.-C. Zhang, Rev. Mod. Phys. {\bf 83},  1057  (2011).

\bibitem{PhysRevB.40.4709}
M. den Nijs and K. Rommelse, Phys. Rev. B {\bf 40},  4709  (1989).

\bibitem{kennedy}
T. Kennedy, Journal of Physics: Condensed Matter {\bf 2},  5737  (1990).

\bibitem{PhysRevLett.77.3443}
M.~A. Mart\'in-Delgado, R. Shankar, and G. Sierra, Phys. Rev. Lett. {\bf 77},
  3443  (1996).

\bibitem{sierra1}
M. Mart\'in-Delgado, J. Dukelsky, and G. Sierra, Physics Letters A {\bf 250},
  430  (1998).

\bibitem{todo}
M. Matsumoto {\it et~al.}, Physica B: Condensed Matter {\bf 329-333},  1010
  (2003).

\bibitem{PhysRevB.76.184428}
J. Almeida, M.~A. Martin-Delgado, and G. Sierra, Phys. Rev. B {\bf 76},  184428
   (2007).

\bibitem{PhysRevB.77.094415}
J. Almeida, M.~A. Martin-Delgado, and G. Sierra, Phys. Rev. B {\bf 77},  094415
   (2008).

\bibitem{1751-8121-41-48-485301}
J. Almeida, M.~A. Martin-Delgado, and G. Sierra, Journal of Physics A:
  Mathematical and Theoretical {\bf 41},  485301  (2008).

\bibitem{kim}
E.~H. Kim, O. Legeza, and J. S\'olyom, Phys. Rev. B {\bf 77},  205121  (2008).

\bibitem{poilblanc}
D. Poilblanc, Phys. Rev. Lett. {\bf 105},  077202  (2010).

\bibitem{chitov}
S.~J. Gibson, R. Meyer, and G.~Y. Chitov, Phys. Rev. B {\bf 83},  104423
  (2011).

\bibitem{Chen}
J. Chen, K.-L. Yao, and L.-J. Ding, Physica A: Statistical Mechanics and its
  Applications {\bf 391},  2306  (2012).

\bibitem{pollmann}
F. Pollmann, E. Berg, A.~M. Turner, and M. Oshikawa, Phys. Rev. B {\bf 85},
  075125  (2012).

\bibitem{hatsugai}
Y. Hatsugai, Journal of the Physical Society of Japan {\bf 75},  123601
  (2006).

\bibitem{hatsugai2}
Y. Hatsugai, New Journal of Physics {\bf 12},  065004  (2010).

\bibitem{affleck_haldane}
I. Affleck and F.~D.~M. Haldane, Phys. Rev. B {\bf 36},  5291  (1987).

\bibitem{hirano_katsura_hatsugai}
T. Hirano, H. Katsura, and Y. Hatsugai, Phys. Rev. B {\bf 77},  094431  (2008).

\bibitem{dagotto_rice}
E. Dagotto and T. Rice, Science {\bf 271},  618  (1996).

\bibitem{nishiyama}
Y. Nishiyama, N. Hatano, and M. Suzuki, Journal of the Physical Society of
  Japan {\bf 64},  1967  (1995).

\bibitem{kim_fath_solyom_scalapino}
E.~H. Kim, G. F\'ath, J. S\'olyom, and D.~J. Scalapino, Phys. Rev. B {\bf 62},
  14965  (2000).

\bibitem{fath_legeza_solyom}
G. F\'ath, O. Legeza, and J. S\'olyom, Phys. Rev. B {\bf 63},  134403  (2001).

\bibitem{todo_munehisa}
S. Todo, M. Matsumoto, C. Yasuda, and H. Takayama, Phys. Rev. B {\bf 64},
  224412  (2001).

\bibitem{chitov2}
G.~Y. Chitov, B.~W. Ramakko, and M. Azzouz, Phys. Rev. B {\bf 77},  224433
  (2008).

\bibitem{berry}
M.~V. Berry, Proceedings of the Royal Society of London. A. Mathematical and
  Physical Sciences {\bf 392},  45  (1984).

\bibitem{white_huse}
S.R.White and D.A.Huse, Phys. Rev. B {\bf 48},  3844  (1993).

\bibitem{PhysRevB.55.2721}
S. Qin, Y.-L. Liu, and L. Yu, Phys. Rev. B {\bf 55},  2721  (1997).

\bibitem{okamoto}
K. Okamoto and K. Nomura, Physics Letters A {\bf 169},  433   (1992).

\bibitem{nomura}
K. Nomura and K. Okamoto, Journal of Physics A: Mathematical and General {\bf
  27},  5773  (1994).

\bibitem{hagiwara}
M. Hagiwara {\it et~al.}, Phys. Rev. Lett. {\bf 65},  3181  (1990).

\bibitem{ng}
T.-K. Ng, Phys. Rev. B {\bf 50},  555  (1994).

\bibitem{lecheminant_orignac}
P. Lecheminant and E. Orignac, Phys. Rev. B {\bf 65},  174406  (2002).

\bibitem{maruyama_hirano_hatsugai}
I. Maruyama, T. Hirano, and Y. Hatsugai, Phys. Rev. B {\bf 79},  115107
  (2009).

\bibitem{hatsugai_BEC_BCS}
M. Arikawa, I. Maruyama, and Y. Hatsugai, Phys. Rev. B {\bf 82},  073105
  (2010).

\bibitem{eggert}
S. Eggert and I. Affleck, Phys. Rev. B {\bf 46},  10866  (1992).

\bibitem{Alps}
A. Albuquerque {\it et~al.}, Journal of Magnetism and Magnetic Materials {\bf
  310},  1187   (2007).

\bibitem{Alps_Bauer}
B. Bauer {\it et~al.}, Journal of Statistical Mechanics: Theory and Experiment
  {\bf 2011},  P05001  (2011).

\bibitem{hirano_katsura_hatsugai_gap}
T. Hirano, H. Katsura, and Y. Hatsugai, Phys. Rev. B {\bf 78},  054431  (2008).

\end{thebibliography}

\end{document}